# On the Dynamics of Kresling Origami Deployment

N. Kidambi and K.W. Wang

*Department of Mechanical Engineering, University of Michigan, Ann Arbor, MI, USA 48109*

**Abstract**

Origami-inspired structures have a rich design space, offering new opportunities for the development of deployable systems that undergo large and complex yet predictable shape transformations. There has been growing interest in such structural systems that can extend uniaxially into tubes and booms. The Kresling origami pattern, which arises from the twist buckling of a thin cylinder and can exhibit multistability, offers great potential for this purpose. However, much remains to be understood regarding the characteristics of Kresling origami deployment. Prior studies have been limited to Kresling structures' kinematics, quasi-static mechanics, or low-amplitude wave responses, while their dynamic behaviors with large shape change during deployment remain unexplored. These dynamics are critical to the system design and control processes, but are complex due to the strong nonlinearity, bistability, and potential for off-axis motions. To advance the state of the art, this research seeks to uncover the deployment dynamics of Kresling structures with various system geometries and operating strategies. A full, six-degree-of-freedom model is developed and employed to provide insight into the axial and off-axis dynamic responses, revealing that the variation of key geometric parameters may lead to regions with qualitatively distinct mechanical responses. Results illustrate the sensitivity of dynamic deployment to changes in initial condition and small variations in geometric design. Further, analyses show how certain geometries and configurations affect the stiffness of various axial and off-axis deformation modes, offering guidance on the design of systems that deploy effectively while mitigating the effects of off-axis disturbances. Overall, the research outcomes suggest the strong potential of Kresling-based designs for deployable systems with robust and tunable performance.

# 1 Introduction

Origami is an ancient paper folding art, employing specific crease patterns to transform a flat sheet into intricate three-dimensional objects. In recent years, origami principles have received significant attention among engineering and science researchers as a way to conceive and analyze new types of structures and materials [1]. This approach has led to a vast array of systems that employ the fundamental kinematic and geometric relationships of foldable systems to achieve dramatic shape-change. Such systems can be folded to minimize their volume for storage and transport, then unfolded or deployed into an extended state in their operational environment [2]. Certain applications, such as deployable space booms and solar arrays [3]–[5], have well-defined storage and operational configurations. These structures are unlikely to undergo a reverse transformation back to the folded state once deployed. Others, such as deployable shelters [6], [7] and self-assembling robots [8]–[10], may require repeated and rapid transformations between states. In both cases, an origami-based approach offers advantages in terms of manufacturability, size, and predictable large-scale shape change [1].

The range of systems that may be designed using origami principles is incredibly vast, but tube-like compositions of origami have been the subject of significant recent interest due to their ability to support loads while offering tunable mechanical response [11]–[13]. Such designs are well-suited for applications that call for uniaxial expansion, such as deployable booms and shelters [14], [15]. Origami tubes can be assembled by stacking sheets with compatible crease patterns to enclose a volume. The Miura pattern [3] is among the most widely-employed crease pattern for this purpose, and has been shown to exhibit features such as large volume change, negative Poisson's ration, and anisotropic stiffness [13], [16]–[19]. Incorporating stiffness elements or fluid pressure to these Miura-based tubes has been shown to enable tunable multistability and energy absorption [12], [20].

Crease patterns for tubular origami structures may also arise from natural phenomena. One example is the Yoshimura pattern [21], whose horizontal valley folds and diagonal mountain folds arise from the axial buckling of a near-ideal thin cylinder. Structures constructed from this pattern and its derivatives have been studied for their mechanical response and energy absorption characteristics [22], [23]. However, they are not suitable for deployable or shape-changing systems, as any deviation from the nominal post-buckled configuration of Yoshimura-patterned tubes causes the panels to experience very large in-plane strains [24], [25]. When a thin-walled cylinder is subject to *twist* buckling, a different type of crease pattern is formed. Commonly known as the Kresling pattern [26], it is characterized by alternating mountain and valley folds angled along the direction of the twist. An example is presented in Figure 1. Like the Yoshimura pattern, the Kresling pattern is not rigidly foldable around its post-buckled configuration [27]. However, unlike the Yoshimura pattern, the Kresling pattern may be bistable, as shown in Figure 1. It may require only moderate panel deformations to compress to a compact state, rendering it far more suitable to the design of deployable tubular structures. Further, while Yoshimura pattern may manifest in many layers of triangulated cells during axial buckling [21], the Kresling pattern only manifests in one layer of triangulated panels for each twist buckling load. This means that multi-layer Kresling origami structures can be folded or assembled manually from individual Kresling modules. Since each Kresling layer may be bistable, such an arrangement can lead to complex, multistable systems in which each constituent module can be independently deployed or collapsed to its extended or compressed state, as shown in Figure 2.

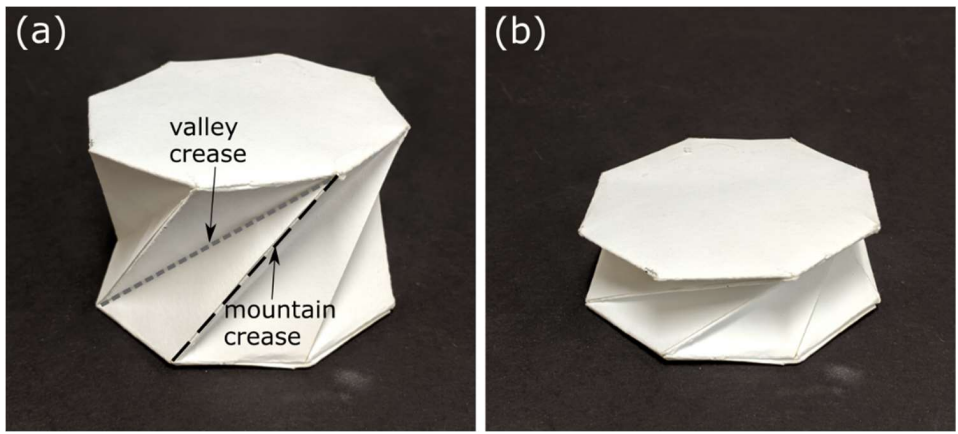

**Figure 1. Kresling origami module showing the arrangement of alternating mountain and valley creases around the circumference, oriented in the direction of twist. This particular geometry exhibits bistability, with (a) an expanded stable state and (b) a compressed stable state.**

Past research on Kresling-inspired structures has revealed that varying geometric parameters can lead to tailorable stiffness and bistability [28], and can bear large loads by exploiting a mechanical diode-like effect to lock into a deployed state [29]. These characteristics, along with the ability to fold into a more compact, flat configuration, make the Kresling pattern an attractive platform for new deployable systems. The prior studies have been limited to kinematics, quasi-static mechanics [28]–[30] or low-amplitude wave propagation [31]. However, deployment is an inherently dynamic process with large-amplitude changes in displacement, and it may occur quickly in Kresling structures due to rapid snap-through motions between stable configurations. These fast dynamics are likely to depend strongly on the structure's geometry. Furthermore, prior investigations have considered only the axial and twist motions of Kresling origami, while the other degrees of freedom and directions have been neglected. In practice, there may be no feasible means to constrain off-axis motions and thus neglecting these motions in analysis may prevent other phenomena from being revealed.

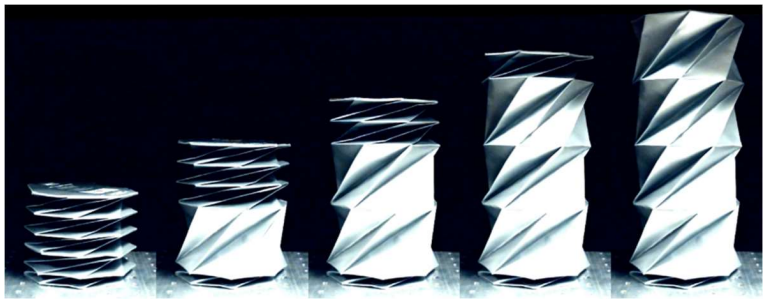

Figure 2. Origami wine tote (BUILT NY, USA) at different stages of deployment. Each configuration shown is stable. The structure is composed of multiple serially-connected Kresling layers.

From the above discussions, while prior research indicates that Kresling origami is well-suited for the design of deployable structures, there are important dynamic features that remain to be understood before this potential can be fully realized. Therefore, to advance the state of the art, the present research objective is to uncover the multi-degree-of-freedom and multi-directional dynamic characteristics of Kresling origami structures during deployment. Through a systematic study of energy landscapes, transient dynamics, and off-axis motions, this investigation seeks to offer insight and guidance for the development of robust and effective deployable Kresling-based systems. To address this research goal, this paper first introduces a full, six-degree-of-freedom model of a Kresling structure using a Newton-Euler approach. This model is then used to study the role of geometric parameters on the structures' mechanics and energy landscapes. Different regions of the energy landscape are explored in further detail, revealing how the multistability and energy barriers between states affects the transient deployment process. Lastly, the response of Kresling structures to off-axis perturbations is discussed, aided by modal analyses of the system at the different stable configurations.

## 2 Model formulation

Unlike many well-studied origami patterns such as the Miura, the Kresling origami pattern is not rigidly foldable and cannot deform from the nominal, expanded state shown in Figure 1(a) purely by folding at the creases [17], [32]. Due to this kinematic incompatibility, transitions between the various configurations shown in Figures 1 and 2 require the triangular panels to bend or stretch to accommodate changes to the crease fold angle. Since traditional models of rigidly foldable origami cannot be employed, past research has adopted different approaches to reflect this non-rigid behavior. One method is to add extra, virtual folds to the triangular facets, allowing them to change shape and deform [32]. This approach provides good insight on the relatively large contribution of panel deformations to the total strain energy in the structure. However, it is ill suited to investigate dynamics since the moving virtual folds are difficult to model using generalized coordinates. Another approach is to treat the creases as bars or trusses that deform axially, resulting in stretching and shearing of the triangular facets [29]. While this approach does not account for Kresling panel bending, it has been shown that panel stretching and shearing are generally sufficient to account for the mechanical response [33]. Furthermore, the treatment of origami creases as truss elements is well suited for dynamic analysis, as the energy potential of each truss is simply a function of the distance between the two nodes to which it connects.

Based on the discussion above, this paper adopts a truss representation of a Kresling module as depicted in Figure 3. To facilitate a parameter study, the model is developed in a nondimensional form. The length scale is defined by the stress-free height $h_0$ of the Kresling module at which all trusses are undeformed. The upper and lower panels are rigid, regular $n$-sided polygons circumscribed by a circle of radius $R_0$.

Since all distances are quantified in terms of $h_0$, a radius $R_0 = 1$ means the circumscribing radius is equal to the module's stress-free height. The panels are connected by *2n* trusses. *Vertical* trusses, denoting the mountain folds of the twist-buckled Kresling crease pattern, connect node $A_i$ on the lower panel with node $B_i$ on the upper panel. *Diagonal* trusses representing the Kresling valley folds connect node $A_i$ with node $B_{i+1}$, for $i \in \{1\ldots n\}$. At the nominal stress-free configuration, the vertical and diagonal trusses have lengths $a_0$ and $b_0$, respectively. Further, there is a *stress-free orientation angle* $\delta_0$, denoting the relative orientation of the upper and lower panels in this configuration. In this model, the polygon sides *n*, radius $R_0$, and stress-free orientation $\delta_0$ are the three nondimensional parameters required to fully define the Kresling module's geometry.

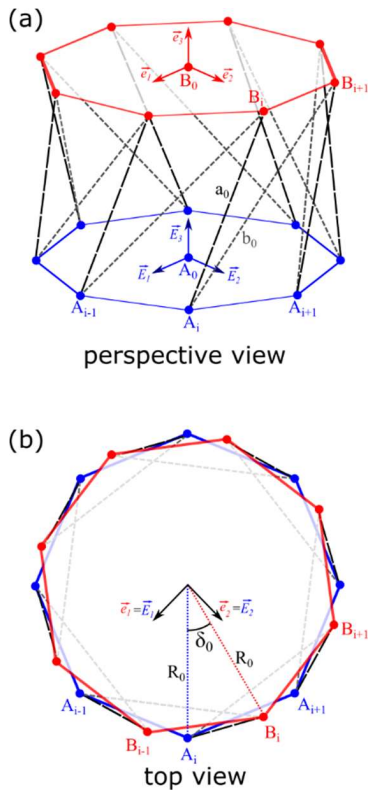

**Figure 3. Truss model of a Kresling module in its stress-free configuration showing (a) a perspective view and (b) a top view. The upper and lower panels are regular, *n*-sided polygons circumscribed by a circle of radius *R₀*. They are connected between vertices by sets of vertical and diagonal creases with unstrained lengths *a₀* and *b₀*, respectively. In the unstressed state, the upper panel is oriented by an angle *δ₀* with respect to the lower panel. Coordinate systems are attached to the two panels. In the stress-free configuration shown, both coordinate systems are equivalent.**

A Newton-Euler approach is adopted to represent system dynamics. Space-fixed orthonormal coordinate vectors $[\vec{E}_1, \vec{E}_2, \vec{E}_3]$ are attached to $A_0$ at the center of the lower panel. Coordinate vectors $[\vec{e}_1, \vec{e}_2, \vec{e}_3]$ are fixed to $B_0$ at the center of the upper panel. At the initial stress-free orientation depicted in Figure 3, the space- and body-fixed basis vectors are identical. The position of $B_0$ with respect to $A_0$ is $\vec{p}_{B_0/A_0} = \vec{E}_3 = \vec{e}_3$ in this configuration, since the stress-free height is used as the length scale for nondimensionalization.

Figure 4 presents a module subject to an arbitrary deformation. This deformation cannot be addressed by prior modelling treatments of Kresling structures [28], [29] as pairs of vertical and diagonal trusses are not identically deformed. In general, $\vec{p}_{B_0/A_0}$ can be written in terms of space-fixed or body-fixed coordinates as:

$$\vec{p}_{B_0/A_0} = p_{A1}\vec{E}_1 + p_{A2}\vec{E}_2 + p_{A3}\vec{E}_3 = p_{B1}\vec{e}_1 + p_{B2}\vec{e}_2 + p_{B3}\vec{e}_3 \tag{1}$$

where $p_{Ai}$ and $p_{Bi}$ are coordinates in frames attached to $A_0$ and $B_0$, respectively. An arbitrary deformation of the module may also impart a rotation of the upper panel, and thus also of the body-fixed coordinates. This rotation is defined by the rotation tensor $\boldsymbol{R}$, as $\vec{e}_i = \boldsymbol{R}\vec{E}_i \quad \forall i \in \{1, 2, 3\}$. The rotation tensor is constructed by employing a standard 3-2-1 set of Euler angles $\Theta = [\gamma, \beta, \alpha]$. They describe any arbitrary 3D rotation as a sequence of three chained rotations around specified axes [34]. The final rotation tensor $\boldsymbol{R}$ is a product of the individual rotation tensors.

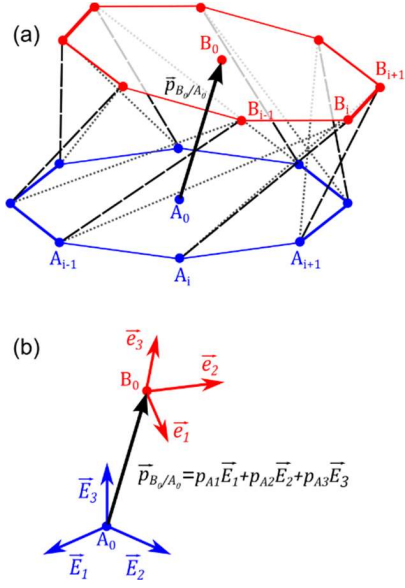

Figure 4. (a) Truss model with an arbitrary displacement and rotation of the upper panel. The center of mass $B_0$ of the upper plate has a position $\vec{p}_{B_0/A_0}$ with respect to the center of mass of the lower panel. The corresponding space- and body-fixed coordinates are presented in (b), showing a rotation of the body-fixed coordinates and a general expression of $\vec{p}_{B_0/A_0}$ in space-fixed coordinates.

$$R = R_1 R_2 R_3 \tag{2}$$

where:

$$R_1 = \begin{bmatrix} \cos\gamma & -\sin\gamma & 0 \\ \sin\gamma & \cos\gamma & 0 \\ 0 & 0 & 1 \end{bmatrix} \tag{3a}$$

$$R_2 = \begin{bmatrix} \cos\beta & 0 & \sin\beta \\ 0 & 1 & 0 \\ -\sin\beta & 0 & \cos\beta \end{bmatrix} \tag{3b}$$

$$R_3 = \begin{bmatrix} 1 & 0 & 0 \\ 0 & \cos\alpha & -\sin\alpha \\ 0 & \sin\alpha & \cos\alpha \end{bmatrix} \tag{3c}$$

Position vectors of the nodes $A_i$ with respect to the center of the lower panel $A_0$ have the following representation in the space-fixed frame:

$$\vec{p}_{A_i/A_0} = R_0 \cos\left(\frac{2\pi}{n}i\right)\vec{E}_1 + R_0 \sin\left(\frac{2\pi}{n}i\right)\vec{E}_2, \qquad \forall i \in \{1...n\} \tag{4}$$

while $\vec{p}_{B_i/B_0}$, the position of nodes on the upper panel with respect to its center of mass, is:

$$\vec{p}_{B_i/B_0} = R_0 \cos\left(\frac{2\pi}{n}i+\delta_0\right)\vec{e}_1 + R_0 \sin\left(\frac{2\pi}{n}i+\delta_0\right)\vec{e}_2 \tag{5a}$$

$$= \left(\mathbf{R}\begin{bmatrix} R_0 \cos\left(\frac{2\pi}{n}i+\delta_0\right) \\ R_0 \sin\left(\frac{2\pi}{n}i+\delta_0\right) \\ 0 \end{bmatrix}\right)^T \begin{bmatrix} \vec{E}_1 \\ \vec{E}_2 \\ \vec{E}_3 \end{bmatrix} \tag{5b}$$

Equations (1), (4), and (5) are combined to write the relative position vectors of the nodes spanned by vertical and diagonal trusses, respectively, as:

$$\vec{p}_{B_i/A_i} = \vec{p}_{B_0/A_0} + \vec{p}_{B_i/B_0} - \vec{p}_{A_i/A_0} \tag{6a}$$

$$\vec{p}_{B_{i+1}/A_i} = \vec{p}_{B_0/A_0} + \vec{p}_{B_{i+1}/B_0} - \vec{p}_{A_i/A_0} \tag{6b}$$

Since the base of the Kresling structure is fixed, the nodes on the lower panel are stationary. Thus, the relative velocity vectors are:

$$\dot{\vec{p}}_{B_i/A_i} = \dot{\vec{p}}_{B_0/A_0} + \dot{\vec{p}}_{B_i/B_0}$$

$$= \dot{p}_{A1}\vec{E}_1 + \dot{p}_{A2}\vec{E}_2 + \dot{p}_{A3}\vec{E}_3 + \vec{\omega}_B \times \vec{p}_{B_i/B_0} \tag{7a}$$

$$\dot{\vec{p}}_{B_{i+1}/A_i} = \dot{\vec{p}}_{B_0/A_0} + \dot{\vec{p}}_{B_{i+1}/B_0}$$

$$= \dot{p}_{A1}\vec{E}_1 + \dot{p}_{A2}\vec{E}_2 + \dot{p}_{A3}\vec{E}_3 + \vec{\omega}_B \times \vec{p}_{B_{i+1}/B_0} \tag{7b}$$

where $\vec{\omega}_B$ is the angular velocity of the upper panel. In general, the basis vectors around which the 3-2-1 Euler angle rotations take place are not orthogonal. Hence, the Euler angle rates $\dot{\boldsymbol{\Theta}} = \begin{bmatrix} \dot{\gamma}, \dot{\beta}, \dot{\alpha} \end{bmatrix}$ are not equivalent to the components of $\vec{\omega}_B$ in any orthonormal basis. Instead, the angular velocity is related to the Euler angle rates through the following transformation:

$$\vec{\omega}_B = \boldsymbol{R}_{\omega_B} \dot{\boldsymbol{\Theta}}^T = \begin{bmatrix} 0 & -\sin\gamma & \cos\beta\cos\gamma \\ 0 & \cos\gamma & \cos\beta\sin\gamma \\ 1 & 0 & \sin\beta \end{bmatrix} \dot{\boldsymbol{\Theta}}^T \qquad (8)$$

Equations (1-8) describe the system's kinematics, but development of governing dynamic equations require a description of system inertias as well as the conservative and nonconservative forces. Inertial properties are defined by assuming the upper and lower panels are circumscribed by disks of radius $R_0$ and thickness $t_0$. The mass of the polygon panels is defined as:

$$m_B = \frac{\rho}{\rho_0} \pi R_0^2 \qquad (9)$$

Where $\frac{\rho}{\rho_0}$ parametrizes the planar mass density and allows the mass to be expressed in terms of other nondimensional parameters. Consequently, the panel's inertia tensor in the body fixed frame is:

$$\boldsymbol{I}_{B_0} = \frac{\rho}{4\rho_0} \pi R_0^4 \begin{bmatrix} 1 & 0 & 0 \\ 0 & 1 & 0 \\ 0 & 0 & 2 \end{bmatrix} \qquad (10)$$

The trusses are modeled as linear elastic elements, parametrized by $r_k = \frac{k_b}{k_a}$, the ratio between the axial stiffness of the diagonal and vertical trusses. Similarly, parameter $r_c = \frac{c_b}{c_a}$ parametrizes the ratio between viscous damping in these trusses. In this research, both ratios are set to be equal to 1 for generality. In practice, however, the relative stiffness and damping of the trusses would depend on the material and manufacturing process. The force exerted at node $B_i$ of the upper panel by the vertical and diagonal trusses is:

$$\vec{F}_{B_i} = \left( \left| \vec{p}_{B_i/A_i} \right| - a_0 \right) \hat{p}_{B_i/A_i} - r_k \left( \left| \vec{p}_{B_i/A_{i-1}} \right| - b_0 \right) \hat{p}_{B_i/A_{i-1}} - 2\zeta \left( \dot{\vec{p}}_{B_i/A_i} \cdot \hat{p}_{B_i/A_i} \right) \hat{p}_{B_i/A_i} - 2\zeta r_c \left( \dot{\vec{p}}_{B_i/A_{i-1}} \cdot \hat{p}_{B_i/A_{i-1}} \right) \hat{p}_{B_i/A_{i-1}} \quad (11)$$

Where the circumflex symbol (^) denotes a unit vector. The torque exerted by the trusses at node $B_i$ around the panel's center of mass $B_0$ is:

$$\vec{T}_{B_i} = \vec{p}_{B_i/B_0} \times \vec{F}_{B_i} \quad (12)$$

The above equations are combined to express the angular and rotational accelerations of the upper panel:

$$\ddot{\vec{p}}_{B_0/A_0} = \frac{1}{m_B} \sum_{i=1}^{n} \vec{F}_{B_i} \quad (13a)$$

$$\dot{\vec{\omega}}_B = \left( \boldsymbol{R}^T \boldsymbol{I}_{B_0} \boldsymbol{R} \right)^{-1} \left( \sum_{i=1}^{n} \vec{T}_{B_i} - \vec{\omega}_B \times \left( \boldsymbol{R}^T \boldsymbol{I}_{B_0} \boldsymbol{R} \right) \vec{\omega}_B \right) \quad (13b)$$

In order to facilitate numerical dynamic analyses, the equations are written in a state-space form:

$$\boldsymbol{x} = \begin{bmatrix} \vec{p}_{B_0/A_0} \\ \boldsymbol{\Theta}^T \\ \dot{\vec{p}}_{B_0/A_0} \\ \vec{\omega}_B \end{bmatrix} \quad (14)$$

and its time derivate is computed as:

$$\dot{\boldsymbol{x}} = \begin{bmatrix} \dot{\vec{p}}_{B_0/A_0} \\ \boldsymbol{R}_{\omega_B}^{-1} \vec{\omega}_B \\ \ddot{\vec{p}}_{B_0/A_0} \\ \dot{\vec{\omega}}_B \end{bmatrix} \quad (15)$$

The governing equations above describe the dynamics of only one Kresling module. However, they can be extended to enable the analysis of multi-module structures by calculating rotation tensors for each panel and computing the displacement and velocity of each of the truss connection points. Given a structure composed of $N$ serially connected modules, the motion of the $j^{th}$ panel can be expressed as:

$$\ddot{\vec{p}}_{j_0} = \frac{1}{m_j} \sum \vec{F}_{j/j+1} + \sum \vec{F}_{j/j-1} \tag{16}$$

$$\dot{\vec{\omega}}_j = \left(\boldsymbol{R}^T I_{j_0} \boldsymbol{R}\right)^{-1} \left(\sum \vec{T}_{j_0/j+1} + \sum \vec{T}_{j_0/j-1} - \vec{\omega}_j \times \left(\boldsymbol{R}^T I_{j_0} \boldsymbol{R}\right) \vec{\omega}_j \right) \tag{17}$$

where $\sum \vec{F}_{j/j+1}$ and $\sum \vec{F}_{j/j-1}$ are the forces on panel $j$ due to truss connections with panel $j+1$ and $j-1$, respectively, while $\sum \vec{T}_{j_0/j+1}$ and $\sum \vec{T}_{j_0/j-1}$ are the torques exerted on the center of mass of panel $j$ due to truss connections with panel $j+1$ and $j-1$, respectively. The base of the structure is fixed to the ground, while the $N^{th}$ panel at the end of the structure only experiences truss forces from connections to the $N$-$1^{st}$ panel. The equations of motion for an $N$-layer structure can be written in state-space form as in Equations (14) and (15). The state vector for an $N$-layer structure has $12N$ entries.

## 3   Quasi-static deployment and energy landscapes

As described in the prior sections, the geometry of Kresling structures can be described by the radius $R_0$, the orientation angle $\delta_0$, and the number of sides to the polygonal panels $n$. Variations in these geometric parameters may yield a range of interesting properties, such as bistability, self-locking, and tunable stiffness [28], [29]. A thorough investigation of the energy landscapes spanned by variations in design parameters and loading conditions will provide insight into the suitability of different designs for deployable structure applications and set the stage for dynamic studies of Kresling deployment.

The model developed in this research does not constrain any degrees of freedom of the system, facilitating investigations of axial and off-axis dynamics. However, to develop initial insight, quasi-static analyses are first performed with a controlled displacement in the $\vec{E}_3$ direction, reflecting the desired deployment direction of Kresling structures [29]. Under quasi-static, pure axial deployment, off-axis motions are not activated, and the structure exhibits a twist-coupled response with displacements and rotations along and around $\vec{E}_3$. An example deployment path is presented in Figure 5(a) for a Kresling module with $R_0 = 0.917$ and $n = 8$. These values reflect measurements from the commercially available wine tote in Figure 1. The stress-free orientation angle is set to $\delta_0 = 32°$ in order to give rise to an asymmetric bistability and a large deployment distance between these stable states, and stiffness ratio is set to $r_k = 1$. Figure 5(a) illustrates how the rotation angle $\gamma$ varies with prescribed Kresling height along a path that minimizes strain energy. As seen from the overlain contour plot, deviations from this deployment path would result in a dramatic increase in strain energy. The strain energy along this path is due to deformations of the vertical and diagonal trusses, as shown in Figure 5(b). When the module is displaced from the stress-free state at $p_{A3} = 1$, the vertical truss strain $\frac{\Delta a}{a_0}$ and diagonal truss strain $\frac{\Delta b}{b_0}$ become nonzero. Figures 5(c) and (d) illustrate how the truss model deforms under axial load for $p_{A3} < 1$, more clearly showing the orientations of the trusses and nodes. This result from the full 6DOF model is consistent with prior results from both the virtual fold models [32] and simplified 2DOF truss models [25], [28].

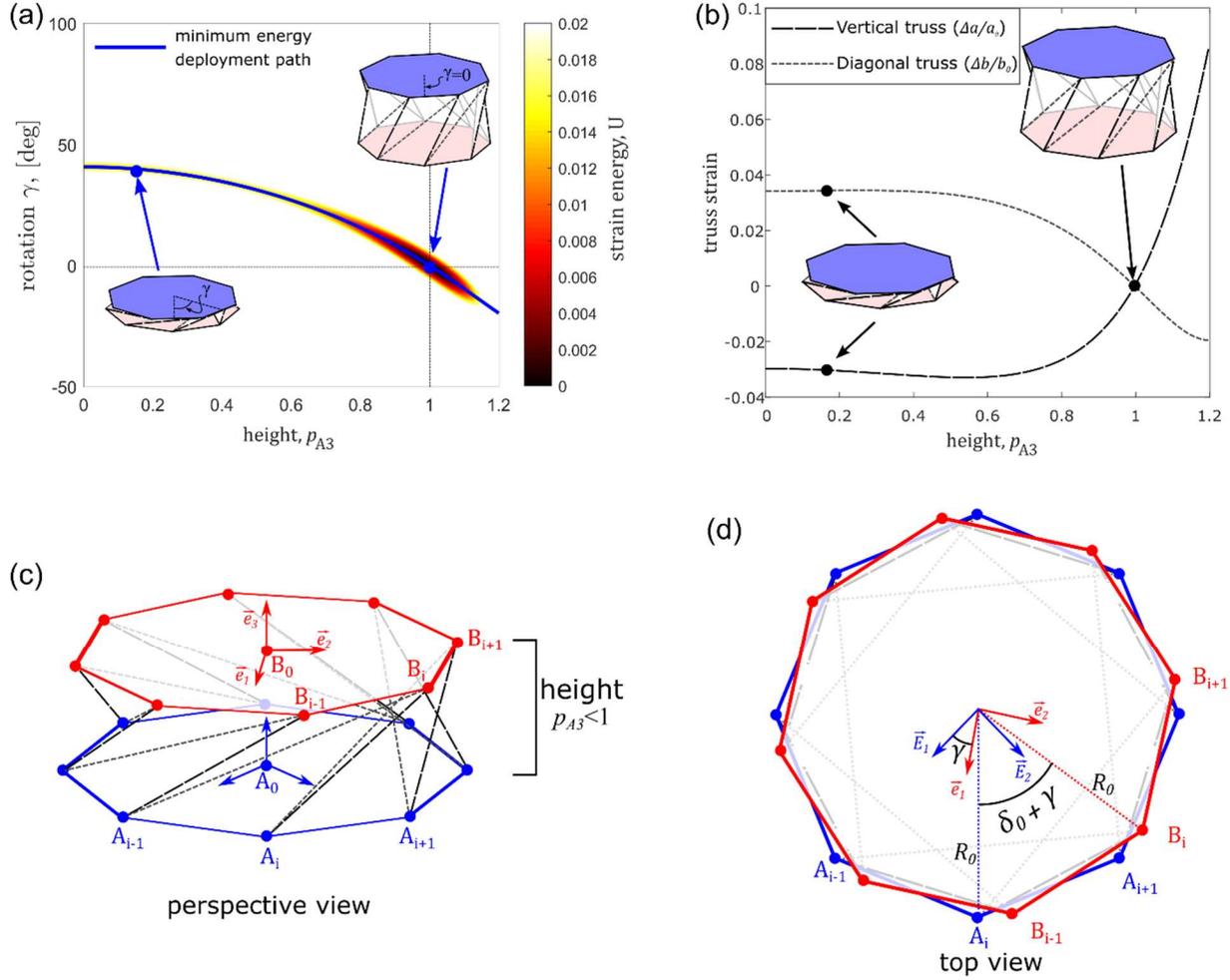

**Figure 5. Under a quasi-static, axial displacement control, the Kresling undergoes a coupled, twist motion where both the rotation angle $\gamma$ and height $p_{A3}$ vary simultaneously along a minimum energy path.** (a) Minimum-energy deployment path along on the ($p_{A3}$, $\gamma$) plane for a Kresling module with stress-free orientation $\delta_0=32°$, $R_0=0.917$, $n=8$, and $r_k=1$. Contour colors indicate strain energy for given $p_{A3}$ and $\gamma$, assuming all other degrees of freedom are fixed. At $p_{A3}=1$, the structure is in its stress-free state with $\gamma=0$. (b) Strains in the vertical and diagonal trusses along the deformation path. (c, d) Perspective and top views of the truss model, respectively, illustrating the rotation $\gamma$ of the upper panel during compression.

### 3.1 Variation of design parameters

To gain more insight into how strain energies vary along this minimum-energy deployment path for different Kresling geometries, Figure 6(a) presents a strain energy landscape for fixed radius $R_0 = 0.917$, truss stiffness ratio $r_k = 1$, and assuming octagonal upper and lower panels $n = 8$. The orientation angle $\delta_0$, a design parameter, is varied along the horizontal axis, and the prescribed Kresling height $p_{A3}$, a loading parameter, is varied along the vertical axis. The red curves denote local minima of strain energy, indicating stable heights $p_{A3}$ for given values of design parameter $\delta_0$. Figure 6(b) shows the theoretical compaction ratio $\eta = h_c/h_d$, where $h_c$ and $h_d$ are the compressed and deployed stable lengths, respectively. The compaction ratio is only relevant for regions of the parameter space that are bistable. There is a range for which the theoretical compaction ratio is 0, though in practice the structure's minimum volume would be constrained by the thickness of the material.

The design space shown in Figure 6 is divided into several different regions based on the qualitative nature of the energy landscape. For region (I) $\delta_0 < 24°$, there is only one stable state at the nominal, stress-free height $p_{A3} = 1$. Kresling modules in this region are *monostable.* Figure 7(a) presents an energy curve along the quasi-static, minimum-energy deployment path of a module in this region, with $\delta_0 = 20°$, clearly illustrating the presence of a single local energy minimum. In region (II) where $24° < \delta_0 < 49.5°$, the fully compressed state is stable, but not stress free. It is therefore referred to as the asymmetrically bistable region. This is exemplified by the example shown in Figure 7(b) for a module with $\delta_0 = 32°$, which shows a local minimum at $p_{A3} = 0$ and a global minimum at $p_{A3} = 1$. A deployment from the fully compressed state to the expanded state would therefore require overcoming an energy barrier. $\delta_0 = 49°$ represents a bifurcation point, above which the fully compressed state is no longer stable. The corresponding strain energy curve is shown in Figure 7(c). In region (III), where $49° < \gamma_0 < 67.5°$, the structure is bistable with one stable state at $p_{A3} = 1$ and a second at $0 < p_{A3} < 1$. Both states are characterized by stress-free trusses with zero strain energy, and the system is thus symmetrically bistable. An example is presented in Figure 7(d) for $\delta_0 = 53°$, which reflects the geometry of the commercially available wine tote depicted in Figure 1. Both stable branches intersect at $p_{A3} = 1$ when $\delta_0 = 67.5°$, leading to a local zero-stiffness property. This geometry is shown in Figure 7(e). For the very large twist angles in region (IV), where $\delta_0 > 67.5°$, the second stable state is at a position $p_{A3} > 1$. As in region (III), the bistability is symmetric with both states having zero strain energy. An example energy curve with $\delta_0 = 80°$ is presented in Figure 7(f). Animations corresponding to all cases in Figure 7 are included in the Supplemental Materials.

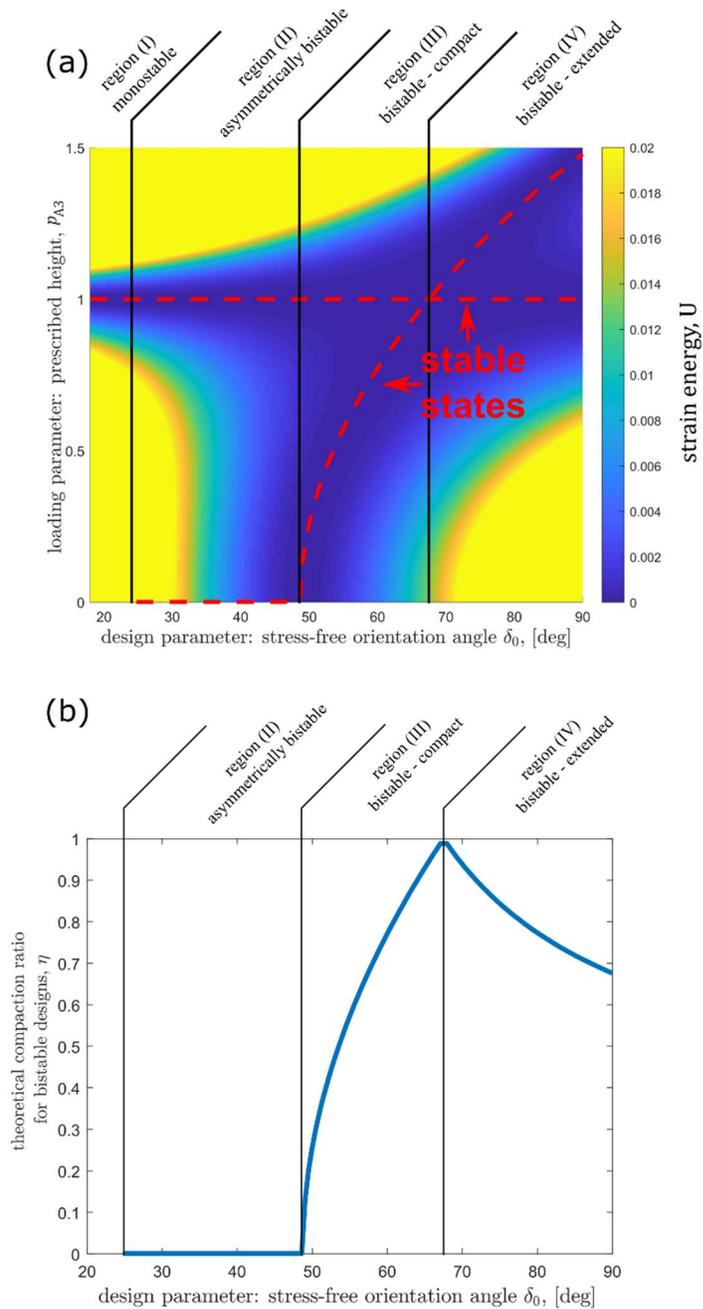

**Figure 6.** (a) Contour plot showing strain energy stored along the minimum energy path, as shown in Figure 5(a), for modules with different values of the stress-free orientation angle $\delta_0$. The other geometric parameters are fixed at *n*=8 and $R_0$=0.917, while stiffness ratio $r_k$=1. Red dashed curves show the stable states, corresponding to local or global minima of strain energy. The presence and location of these states leads to a natural division of the design space into four regions with different stability characteristics. (b) Compaction ratio $\eta$ between the compressed and extended stable lengths as the design parameter is varied.

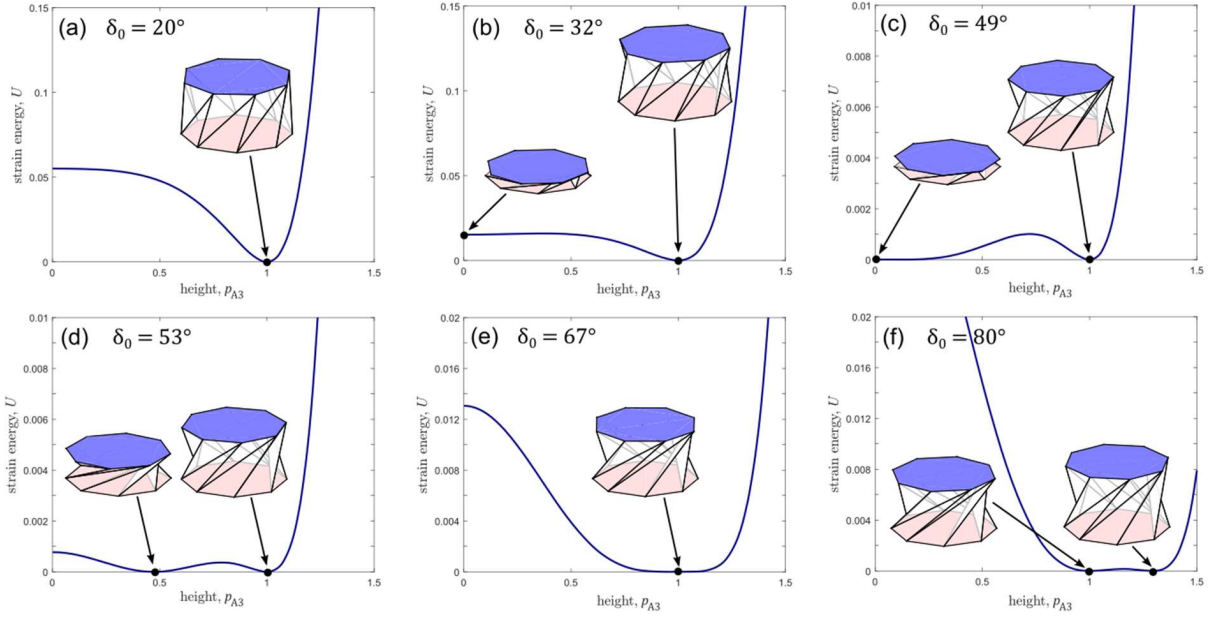

**Figure 7. Strain energy landscapes along minimum energy deployment paths for Kresling modules with varying stress-free orientations $\delta_0$.** (a) A module with $\delta_0$=20° lies in the monostable region (I) in Figure 6, and has just one stable position at its stress-free height of 1. (b) When $\delta_0$=32°, the system is in the asymmetric bistable region (II). There is one global energy minimum at the stress-free height and a local minimum at $p_{A3}$=0. (c) For $\delta_0$=49°, the configuration at $p_{A3}$=0 is a second global energy minimum and corresponds to the bifurcation seen in Figure 6 marking the transition from region (II) to region (III). (d) For $\delta_0$=53° in region (III), the second stable position is at a compact state $p_{A3}$<1 while for (e) $\delta_0$=67°, there is a double root at the stress-free height, resulting in locally-zero-stiffness around the stable point. (f) For $\delta_0$=80°, the second energy minimum is at a further extended state $p_{A3}$>1, placing this module in region (IV) in Figure 6. Corresponding animations are in the Supplemental Materials.

A similar parameter analysis is conducted for $R_0$, the radius of the Kresling module. A contour plot summarizing the results is presented in Figure 8(a). The horizontal axis denotes variations in the design parameter $R_0$, for which a few examples are visualized in Figure 8(b). The vertical axis denotes the prescribed module height. The other geometric parameters are fixed at $\delta_0 = 53°$ and $n = 8$. For small radii $R_0 < 1.02$, the modules are bistable. They have a stable nominal stress-free height at $p_{A3} = 1$ and another stable state at $p_{A3} < 1$. The potential energy landscapes are qualitatively similar to region (III) in Figure 6. For larger radii $1.02 < R_0 < 1.42$, the system is asymmetrically bistable. The stable, fully compact state at $p_{A3} = 0$ has some nonzero strain energy, similar to region (II) in Figure 6. For very large radii $R_0 > 1.42$

, the fully compact state loses stability and the system is simply monostable. Figure 8(c) shows the theoretical compaction ratio $\eta$ for designs with $R_0$ in the bistable and asymmetrically bistable regions.

The results of Figures 6 and 8 illustrate a few key points regarding the suitability of various designs for deployable structure applications. Structures in regions (I) and (IV) in Figure 6 have large strain energy in the compact state, and it may thus be infeasible to fully compact and constrain them without a sufficiently large external force. On the other hand, once released from the compacted state, structures in these regions will automatically deploy a stress-free state $p_{A3}=1$ (or $p_{A3} \geq 1$ in the case of region (IV)) without the need to overcome an additional energy barrier. If practical packaging restrictions permit, monostability without exceedingly high-energy compact configurations may be achieved by increasing the radius as shown in Figure 8. Structures in regions (II) and (III) are stable for some compacted state $p_{A3}<1$, so they can be collapsed and stored without the need for an extra constraining force. However, as illustrated in Figures 7(b-d), transitions from the compacted to the expanded stress-free state require overcoming an energy barrier. The highly nonlinear nature of the dynamic response of bistable and multistable systems means that predicting the final configuration from initial conditions is not trivial [35], [36].

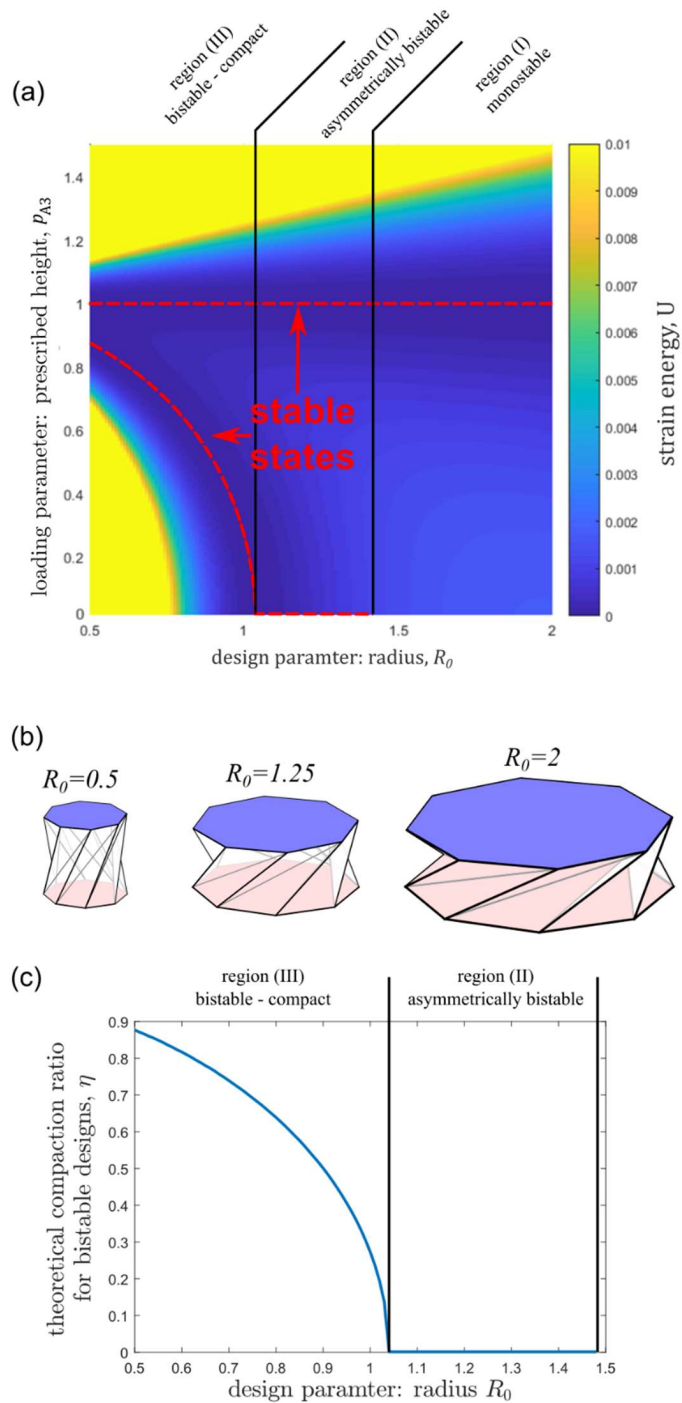

**Figure 8.** (a) Contour plot showing strain energy stored along the minimum energy path for modules with different values of the stress-free orientation angle $R_0$. The other geometric parameters are fixed at $n=8$ and $\delta_0=53°$, while stiffness ratio $r_k=1$. Red dashed curves show the stable states, corresponding to local or global minima of strain energy. The presence and location of these states leads to a natural division of the design space into three regions with different stability characteristics, with qualitatively similar characteristics to equivalent regions in Figure 6. (b) Images showing the effect of $R_0$ on the shape of the stress-free state at $p_{A3}=1$. (c) Theoretical compaction ratio $\eta$ for designs in the bistable regions.

## 4 Dynamic analysis

The quasi-static analyses presented in the prior section shed light on the various qualitative mechanical properties and stability characteristics of Kresling structures. However, they are insufficient to properly understand the dynamics of Kresling deployment. Often, origami-inspired structures are intended to reconfigure quickly [6], [32], and rapid shape change may not smoothly follow the minimum energy paths presented in Figure 5 and 7 in structures with multiple degrees of freedom [37]. Furthermore, quasi-static analyses were conducted only along the $\vec{E}_3$ axis, which is the direction in which Kresling structures are designed to deploy [15]. However, perturbations, manufacturing imperfections, and transverse loads may excite off-axis dynamics as well. To address these points, this section discusses the dynamic responses of Kresling structures during deployment. Simulations are conducted using MATLAB's ODE45 solver, which is a fifth order Runge-Kutta method. In these dynamic analyses, the planar mass density of each plate is assigned as $\frac{\rho}{\rho_0} = \frac{1}{\pi R_0^2}$. The damping ratio $\zeta$ is selected such that a module supported only by vertical trusses is critically damped and $r_c = 1$. Unless otherwise noted, geometric parameters are $R_0 = 0.917$, and $n = 8$, while the stiffness ratio is $r_k = 1$. Boundary conditions reflect free deployment from a fixed base. The base of the structure is attached to the ground and the deployment process is powered by stored strain energy when the structure is packaged and under compression. During and after deployment, no loads of constraints are applied to the end of the structure.

### 4.1 Axial deployment

#### 4.1.1 Deployment of a four-module structure

As described in a prior section, the different stability regions spanned by the variations of $\delta_0$ and $R_0$ shown in Figures 6 and 8 may require different approaches and strategies for system deployment. For example, deployment from any compact or compressed state to the extended, stress-free configuration is inevitable in the monostable region (I). There is no other energy minimum to which the system may eventually settle. In this parameter region, the Kresling structure may be constrained and transported in a compressed configuration. When the constraint is released, the system will naturally expand and release the stored elastic energy, settling in the final deployed state. The results in Figure 9 present an example for four serially connected modules with $R_0 = 0.917$, $n = 8$, $\delta_0 = 22°$, and $r_k = 1$. The snapshots presented above show the initial, intermediate, and final states of the structure. The second snapshot illustrates that the deployment does not necessarily occur simultaneously or sequentially. This is due to the strong geometric nonlinearity and influence of the reaction forces at the fixed boundary [35]. The third snapshot shows a small overshoot, which is reasonable given the selection of damping ratio $\zeta$.

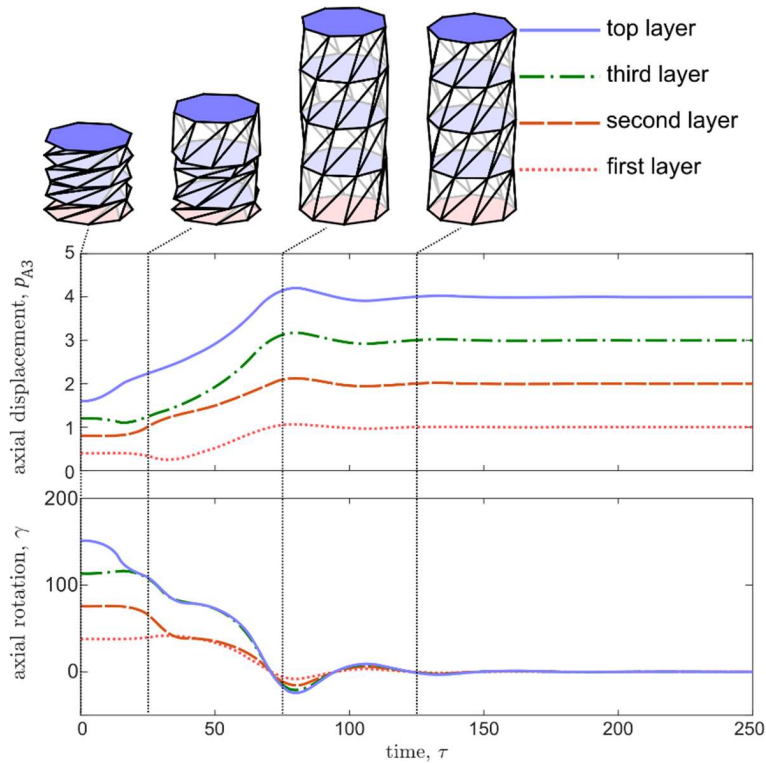

**Figure 9.** (a) Position $p_{A3}$ along $\vec{E}_3$ and first Euler rotation angle $\gamma$ during dynamic deployment of a four-layer chain of Kresling modules with $\delta_0$=22°. The structure is fixed at its base, compressed to 40% of its initial height, and then released. It settles in its extended, stress-free state. Snapshots before, during, and after deployment are shown above indicate a small overshoot.

Kresling structures in the bistable regions of Figures 6 and 8 may have less predictable dynamic responses, since there are two states to which the system may settle. For example, Figure 10 considers the deployment of a four-module structure with $\delta_0 = 55°$ initially compressed to (a) 22.5% and (b) 25% of its extended, stress-free height. The systems are then released and allowed to come to rest. The initial condition in (a) causes all four modules to deploy to the fully extended configuration. On the other hand, the slightly lower strain energy in the initial condition in (b) means that one of the modules does not cross the energy barrier and deploy to the extended state. Figure 10(c) presents a case with 22.5% initial compression, but with damping reduced by half. The results show that the reduced dissipation results in residual kinetic energy after all modules are deployed. This residual energy results in a larger overshoot and causes the first layer to cross back over its energy barrier and to compressed stable configuration. The results in Figure 10 illustrate the sensitivity of Kresling deployment to changes in initial conditions and damping. This sensitivity is expected given the highly nonlinear, multistable nature of the system, but necessitates careful selection of initial conditions to achieve desired deployment performance.

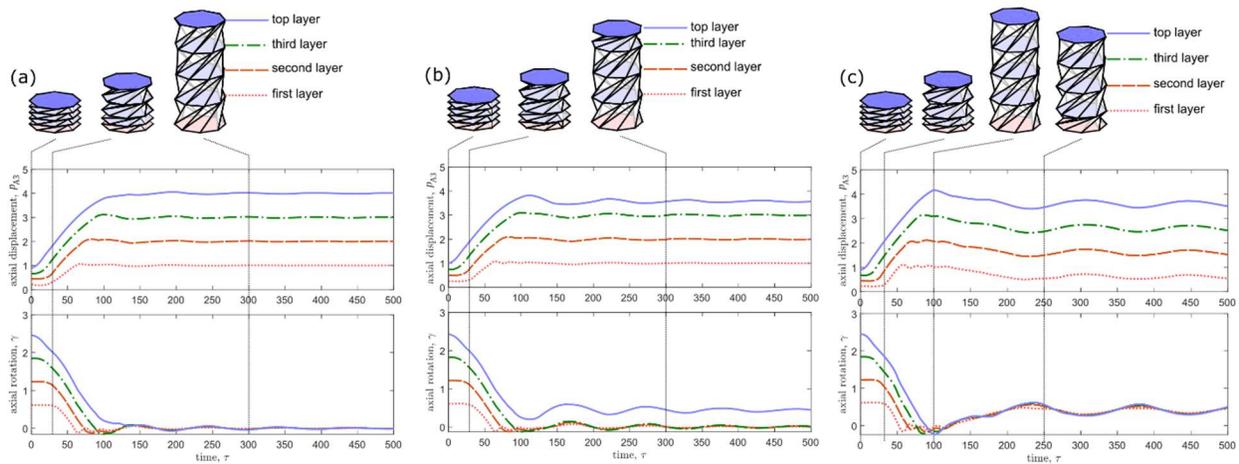

**Figure 10. Position $p_{A3}$ along $\vec{E}_3$ and first Euler rotation angle $\gamma$ during dynamic deployment of a four-module system with $\delta_0$=55°, placing the structure in the bistable region (II) in Figure 6. Responses when the structure is compressed to (a) 22.5% and (b) 25% of its nominal, stress-free length and then released. The slightly larger quantity of stored energy in (a) is sufficient to cause all modules to overcome the energy barrier between bistable states and deploy to the extended configuration. Snapshots of the deployment process are shown above. (c) An underdamped case where the linearized damping ratio is reduced by half compared with (a) and (b) and the structure is initially compressed to 22.5% of its nominal length. The response shows an overshoot due to residual kinetic energy after all layers are deployed, which causes the first layer to overcome the energy barrier in the reverse direction and return back to a compressed stable state.**

*4.1.2  Basins of attraction of a single module*

To gain further insight into how initial conditions and geometries may affect transient deployment, Figure 11 shows a basin of attraction map for a single module in a portion of the bistable regions in Figure 6 and 8. The horizontal axis denotes variations in the design parameter $\delta_0$ or $R_0$, while the vertical axis indicates the initial compression as a fraction of the nominal, stress-free height. Colors indicate the final configuration for a given design and initial condition. To achieve reasonable fidelity, $\delta_0$ and $R_0$ are varied in increments of 0.02° and 0.01, respectively. Since the governing equations are deterministic, the simulations are repeatable for a given combination of design and initial condition. Dark colored squares indicate that the module comes to rest in its expanded state where $p_{A3} = 1$, while lighter squares indicate that the module comes to rest at some stable state $p_{A3} < 1$. Red dashed lines denote the positions of these stable states.

While Figure 11 aggregates dynamic results for structures composed of just one module, it nevertheless shows how the transient deployment process may be highly sensitive to variations in design and initial compression. From a practical standpoint, it may be prudent to design structures and specify initial conditions that lie in regions that are less sensitive to changes in these parameters. This would help ensure predictable deployment performance that is less likely to be compromised by variability in manufacturing and/or initial conditions.

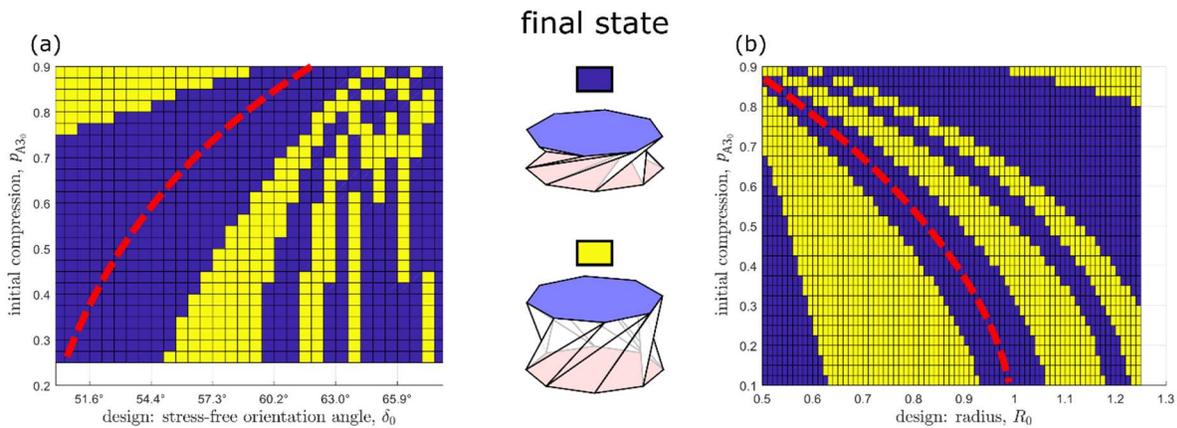

**Figure 11. Basin of attraction maps showing aggregated results of dynamic simulations starting at different values of initial compression, $p_{A30}$ as a design parameter is varied along the horizontal axis. (a) Horizontal axis varies the stress-free orientation angles $\delta_0$ within the bistable region (II) in Figure 6. (b) Variations of the radius $R_0$ within the bistable region**

(II) in **Figure 8**. The red dashed curve shows the locations of the stable states of $p_{A3}$ for the given value of the design parameter. Blue regions indicate that the final configuration is the compressed stable state, while yellow regions denote final states that are in the stress-free, expanded states.

*4.2  Off-axis response*

The results and analyses presented in the prior sections have been pursued on models that include all six degrees of freedom, although only axial motions were perturbed when varying initial conditions. As a result, only dynamic responses in $p_{A3}$ and $\gamma$ were observed. In practical applications, off-axis motions may be perturbed for a variety of reasons, including imperfections in fabrication, the influence of gravitational and other forces, or disturbances from the environment. In order to provide some initial insight into the performance of Kresling structures to off-axis perturbations, Figure 12 presents three examples of transient responses for a structure composed of four modules with $\delta_0 = 55°$ subject to some off-axis perturbation. Curves trace the three components of displacement and the three Euler angles of the upper panel in the chain. In Figure 12(a) the last panel is given an initial velocity in the vertical direction: $\dot{p}_{A3} = 0.1$; and initial angular velocity along the off-axis $\vec{E}_2$ direction: $\omega_{B2_0} = 0.3$. The transient response shows that the system deploys to its fully extended state, and that the off-axis oscillations in $p_{A1}$, $p_{A2}$, $\alpha$, and $\beta$ diminish rather quickly. Figure 12(b) presents a response with initial angular velocity component $\omega_{B2_0} = 0.4$. Under this initial condition, the system does not fully deploy, and three of the four modules settle in the compressed stable state. Moreover, the off-axis oscillations take much longer to diminish than in 12(a). Figure 12(c) presents a case where there are no off-axis perturbations due to initial conditions, but where the two vertical trusses along one side of the structure have stiffness and damping reduced by 50%. This may reflect a manufacturing defects or damage due to impact or wear. The initial compression is the same as in Figure 10(a), but imperfection of the trusses on one side prevents full deployment and causes off-axis oscillations. Images above all three plots in Figure 12 show snapshots at specified points in time. Off-axis oscillations in Figure 12(a) are smaller and diminish more quickly than in Figure 12(b) and (c), suggesting that the fully deployed, stress-free configurations may be more robust to off-axis perturbations.

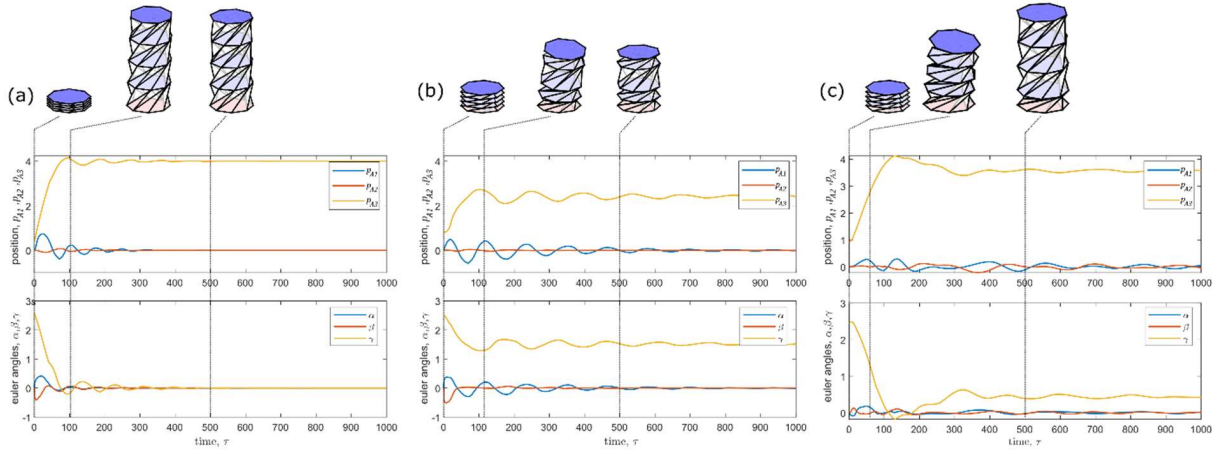

**Figure 12. Transient response of a module with $\delta_0$=55° from an initially compressed state with (a) initial velocities ($\dot{p}_{A3}=0.1; \omega_{B2_0}=0.3$). All modules in the system fully deploy to the extended, stress-free configuration and the off-axis perturbation is quickly diminished. (b) An initial condition of $\omega_{B2_0}=0.4$ results in only one of the four modules deploying to the fully extended state, and the transient response shows significant oscillation in the axial and off-axis directions. (c) Response with no perturbations in the off axis direction, but where trusses along one side of the structure have stiffness and damping reduced by 50%. Overall, the results suggest that the extended state is significantly stiffer and more robust to off-axis perturbations than the compressed state.**

To further understand the off-axis responses in the expanded and compressed stable states, this section analyzes the structures' vibration modes in both configurations. The system is first linearized around the stable states and small amplitude motions are assumed. This assumption means that Euler angle rates $\dot{\Theta}$ and angular velocity components of $\vec{\omega}_B$ in the $\vec{E}_3$ coordinate basis are approximately equivalent. The mass matrix and stiffness matrix are constructed as follows:

$$M = \begin{bmatrix} m_B I_3 & 0 \\ 0 & I_{B_0} \end{bmatrix} \tag{18a}$$

$$K = \begin{bmatrix} \dfrac{\partial \vec{F}_B}{\partial \vec{r}_{B_0/A_0}} & \dfrac{\partial \vec{F}_B}{\partial \Theta} \\ \dfrac{\partial \vec{T}_B}{\partial \vec{r}_{B_0/A_0}} & \dfrac{\partial \vec{T}_B}{\partial \Theta} \end{bmatrix} \tag{18b}$$

where $I_3$ is the 3x3 identity matrix and the components of the stiffness matrix $K$ are calculated by taking partial derivatives of equations (11) and (12).

The resulting eigenvalue problem is solved in Mathematica to yield natural frequencies and mode shapes. This procedure is carried out for both stable branches shown in Figure 6 for variations of the design parameter $\delta_0$. Considering only the three lowest eigenvalues corresponding to the three lowest natural frequencies, the results reveal a decoupling between axial and off-axis modes. That is, the eigenvectors either lie on the subspace spanned by $(p_{A3}, \gamma)$, indicating a pure axial mode, or are orthogonal to it. This is visually interpreted in Figure 13(a). The axial mode has components only in $p_{A3}$ and $\gamma$, denoting motion along and around the $\vec{E}_3$ axis. The off-axis modes have components in the $p_{A1}$, $p_{A2}$, $\alpha$, and $\beta$ directions and are fully uncoupled from axial motions.

Figure 13(b) shows the modal frequencies of the axial and off-axis modes of the linearized stable states in Figure 6 as the parameter $\delta_0$ is varied. For reference, the inset shows the contour plot of Figure 6. Stable branch A corresponds to the nominal, stress-free stable configuration at $p_{A3} = 1$ and branch B denotes the compressed or extended stable states in the bistable region. Due to symmetry, the two lowest-frequency off-axis modes have the same eigenfrequency but orthogonal eigenvectors. Aside from a small portion of branch A for $\delta_0 < 28°$, the axial mode is the fundamental mode for the majority of the design space. Furthermore, for $\delta_0 < 67°$, the axial and off axis modes in branch A have higher model frequencies than their counterparts in branch B. These findings help explain why the results of Figure 12 showed more robustness to off-axis perturbations when fully deployed, and can help guide the design of deployable Kresling structures that are relatively soft in the axial direction and thus easy to deploy, but are stiff in the off-axis direction and are thus robust to perturbations. Another interesting outcome in Figure 13(b) is the clear zero-stiffness axial mode observed for $\delta_0 = 67°$. Kresling structures designed near this point may have attractive vibration isolation properties.

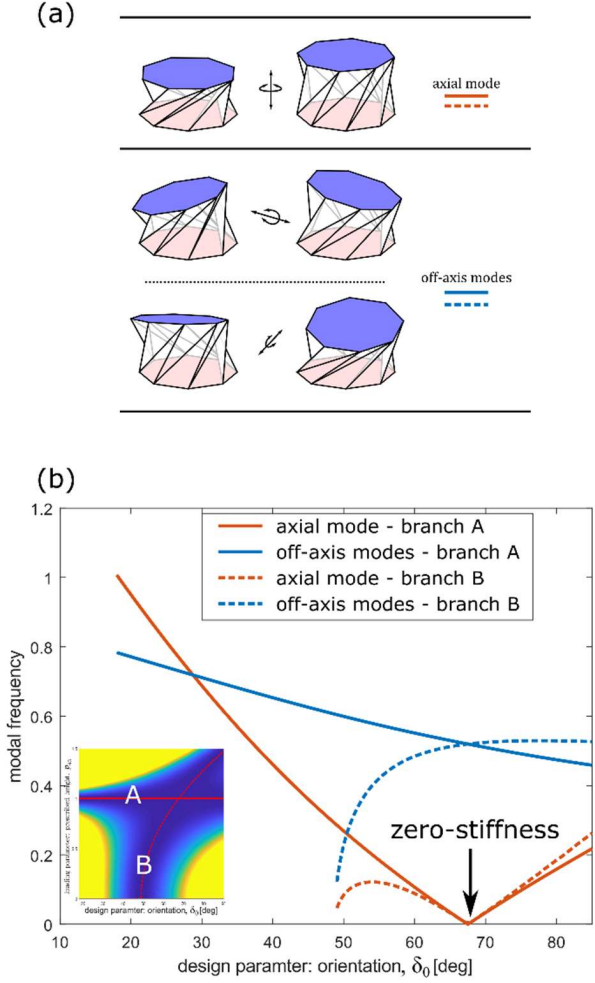

**Figure 13.** (a) Images illustrating the shapes of the first three linearized modes around the nominal, stress-free configuration. The axial and off-axis modes are decoupled. (b) Modal frequencies of the axial and off-axis modes for different values of the stress-free orientation angle $\delta_0$. Other parameters are fixed at $R_0$=0.917, $n$=8, and $r_k$=1. Branch A corresponds to the nominal stress-free state at $p_{A3}$=1, while branch B corresponds to the compressed or extended stable configuration in the bistable region of Figure 6, which is presented as an inset for reference. The axial mode is generally the fundamental mode and has a zero-stiffness response at $\delta_0$=67°.

## 5 Conclusions

This research explores the rich mechanical properties and deployment dynamics of Kresling origami-inspired structures. Through a systematic study of energy landscapes, transient dynamics, and off-axis motions, this investigation offers new insight to the potential for Kresling origami as a platform to develop deployable systems. To capture dynamic responses in all six degrees of freedom, this paper develops a truss model that accounts for off-axis motions that have often been overlooked in prior study. Systematic quasi-static analyses are conducted on Kresling structures with varying geometric properties. It is shown that by tuning these geometric parameters, the energy landscapes of the Kresling module may qualitatively change between monostability, asymmetric bistability, and symmetric bistability. Each region of the design space may be best suited for a different deployment strategy. For example, monostable structures can be deployed simply by compressing the structure to a compact state, then releasing the constraining force when needed. However, monostable Kresling structures may have large strain energy when compacted, and applying sufficient constraining force may pose practical challenges. The deployment of systems in the bistable region may not require as much energy, but deployment to the full, extended state is sensitive to small changes in initial conditions and geometric parameters. Basin of attraction maps help interpret aggregate transient dynamic responses, providing insight into the Kresling designs that may limit the negative effects of this sensitivity. Further analysis of the dynamic response of Kresling structures reveals that certain designs may be robust to perturbations in the off-axis directions in the fully extended state, but sensitive to such perturbations in a compressed configuration. Linearized modal analyses in the extended and compressed state gives insights into this behavior and helps guide the designs that balance deployability in the axial direction with robustness to out-of-axis loads and perturbations. Overall, this research provides good understanding on the rich mechanical properties and dynamic responses of Kresling structures during the deployment process, offering new potential for the development of robust and effective deployable structures.

## Acknowledgements

This research is supported by the National Science Foundation under grant number 1634545 and by the Ford Motor Company under the Ford/U-M Innovation Alliance Program.

## References

[1] A. Lebée, "From folds to structures, a review," *Int. J. Sp. Struct.*, vol. 30, no. 2, pp. 55–74, 2015.

[2]  J. Morgan, S. P. Magleby, and L. L. Howell, "An Approach to Designing Origami-Adapted Aerospace Mechanisms," *J. Mech. Des.*, vol. 138, no. 5, p. 052301, 2016.

[3]  K. Miura, "Method of packaging and deployment of large membranes in space," in *31st Congress of the International Astronautical Federation*, 1985.

[4]  H. M. Y. C. Mallikarachchi and S. Pellegrino, "Deployment dynamics of ultrathin composite booms with tape-spring hinges," *J. Spacecr. Rockets*, vol. 51, no. 2, pp. 604–613, 2014.

[5]  L. Puig, A. Barton, and N. Rando, "A review on large deployable structures for astrophysics missions," *Acta Astronaut.*, vol. 67, no. 1–2, pp. 12–26, 2010.

[6]  A. P. Thrall and C. P. Quaglia, "Accordion shelters: A historical review of origami-like deployable shelters developed by the US military," *Eng. Struct.*, vol. 59, pp. 686–692, Feb. 2014.

[7]  J. Cai, X. Deng, Y. Xu, and J. Feng, "Geometry and motion analysis of origami-Based Deployable Shelter structures," *J. Struct. Eng.*, vol. 141, no. 10, p. 06015001, 2015.

[8]  S. Felton, M. Tolley, E. D. Demaine, D. Rus, and R. Wood, "A method for building self-folding machines," *Science (80-. ).*, vol. 345, no. 6197, pp. 644–646, Aug. 2014.

[9]  W. Gao, K. Huo, J. S. Seehra, K. Ramani, and R. J. Cipra, "HexaMorph: A reconfigurable and foldable hexapod robot inspired by origami," *IEEE Int. Conf. Intell. Robot. Syst.*, no. Iros, pp. 4598–4604, 2014.

[10] S. Miyashita, S. Guitron, S. Li, and D. Rus, "Robotic metamorphosis by origami exoskeletons," *Sci. Robot.*, vol. 2, no. 10, pp. 1–7, 2017.

[11] K. C. Cheung, T. Tachi, S. Calisch, and K. Miura, "Origami interleaved tube cellular materials," *Smart Mater. Struct.*, vol. 23, no. 9, 2014.

[12] S. Li and K. W. Wang, "Fluidic origami with embedded pressure dependent multi-stability: a plant inspired innovation," *J. R. Soc. Interface*, vol. 12, no. 111, p. 20150639, 2015.

[13] E. T. Filipov, T. Tachi, and G. H. Paulino, "Origami tubes assembled into stiff, yet reconfigurable structures and metamaterials," *Proc. Natl. Acad. Sci.*, vol. 112, no. 40, pp. 12321–12326, 2015.

[14] M. C. Natori, N. Katsumata, H. Yamakawa, H. Sakamoto, and N. Kishimoto, "Conceptual model study using origami for membrane space structures - a perspective of origami-based engineering," in *IDETC/CIE 2013*, 2013.

[15] L. Wilson, S. Pellegrino, and R. Danner, "Origami sunshield concepts for space telescopes," *54th AIAA/ASME/ASCE/AHS/ASC Struct. Struct. Dyn. Mater. Conf.*, 2013.

[16] S. Kamrava, D. Mousanezhad, H. Ebrahimi, R. Ghosh, and A. Vaziri, "Origami-based cellular metamaterial with auxetic, bistable, and self-locking properties," *Sci. Rep.*, vol. 7, p. 46046, 2017.

[17] E. T. Filipov, K. Liu, T. Tachi, M. Schenk, and G. H. Paulino, "Bar and hinge models for scalable analysis of origami," *Int. J. Solids Struct.*, vol. 124, pp. 26–45, 2017.

[18] H. Fang, S. C. A. Chu, Y. Xia, and K. W. Wang, "Programmable Self-Locking Origami Mechanical Metamaterials," *Adv. Mater.*, vol. 30, no. 15, pp. 1–9, 2018.

[19] H. Yasuda and J. Yang, "Reentrant origami-based metamaterials with negative Poisson's ratio and bistability," *Phys. Rev. Lett.*, vol. 114, p. 185502, 2015.

[20] H. Fang, S. Li, and K. W. Wang, "Self-locking degree-4 vertex origami structures," *Proc. R. Soc. A Math. Phys. Eng. Sci.*, vol. 472, no. 2195, 2016.

[21] Y. Yoshimura, "Mechanism of buckling of a circular cylindrical shell under axial compression," *NASA TM 1390*, pp. 1–45, 1955.

[22] S. D. Guest and S. Pellegrino, "The Folding of Triangulated Cylinders, Part I: Geometric Considerations," *J. Appl. Mech.*, vol. 61, no. 4, p. 773, 1994.

[23] S. D. Guest and S. Pellegrino, "The Folding of Triangulated Cylinders, Part III: Experiments," *J. Appl. Mech.*, vol. 63, no. 1, p. 77, 1996.

[24] K. Miura and T. Tachi, "Synthesis of rigid-foldable cylindrical polyhedra," *J. ISIS-Symmetry, Spec. Issues Festival-Congress Gmuend, Austria*, pp. 204–213, 2010.

[25] J. Cai, X. Deng, Y. Zhou, J. Feng, and Y. Tu, "Bistable behavior of the cylindrical origami structure with Kresling pattern," *J. Mech. Des.*, vol. 137, no. 6, p. 061406, 2015.

[26] B. Kresling, "Origami-structures in nature: lessons in designing 'smart' materials," in *Materials Research Society Symposium Proceedings*, 2012, vol. 1420.

[27] F. Bös, M. Wardetzky, E. Vouga, and O. Gottesman, "On the Incompressibility of Cylindrical Origami Patterns," *J. Mech. Des. Trans. ASME*, vol. 139, no. 2, pp. 1–9, 2017.

[28] H. Yasuda, T. Tachi, M. Lee, and J. Yang, "Origami-based tunable truss structures for non-volatile mechanical memory operation," *Nat. Commun.*, vol. 8, no. 1, pp. 1–6, 2017.

[29] Z. Zhai, Y. Wang, and H. Jiang, "Origami-inspired, on-demand deployable and collapsible mechanical metamaterials with tunable stiffness.," *Proc. Natl. Acad. Sci. U. S. A.*, vol. 0, p. 201720171, 2018.

[30] J. Cai, Y. Liu, R. Ma, J. Feng, and Y. Zhou, "Nonrigidly Foldability Analysis of Kresling Cylindrical Origami," *J. Mech. Robot.*, vol. 9, no. 4, p. 041018, 2017.

[31] H. Yasuda, Y. Miyazawa, E. G. Charalampidis, C. Chong, P. G. Kevrekidis, and J. Yang, "Origami-based impact mitigation via rarefaction solitary wave creation," *arXiv Prepr.*, pp. 1–19, 2018.

[32] A. Pagano, T. Yan, B. Chien, A. Wissa, and S. Tawfick, "A crawling robot driven by multi-stable origami," *Smart Mater. Struct.*, vol. 26, no. 9, 2017.

[33] K. Liu and G. H. Paulino, "Nonlinear mechanics of non-rigid origami : an efficient computational approach," *Proc. R. Soc. A Math. Phys. Eng. Sci.*, vol. 473, p. 20170348, 2017.

[34] O. M. O'Reilly, *Intermediate dynamics for engineers: A unified treatment of Newton-Euler and Lagrangian mechanics*. Cambridge, UK: Cambridge University Press, 2008.

[35] N. Kidambi, R. L. Harne, and K. W. Wang, "Energy capture and storage in asymmetrically multistable modular structures inspired by skeletal muscle," *Smart Mater. Struct.*, vol. 26, no. 8, p. 085011, 2017.

[36] C. Zhang, R. L. Harne, B. Li, and K. W. Wang, "Reconstructing the transient, dissipative dynamics of a bistable Duffing oscillator with an enhanced averaging method and Jacobian elliptic functions," *Int. J. Non. Linear. Mech.*, vol. 79, pp. 26–37, 2016.

[37] C. Liu and S. M. Felton, "Transformation Dynamics in Origami," *Phys. Rev. Lett.*, vol. 121, no. 25, p. 254101, 2018.